\documentclass{PoS}
\usepackage{bm}

\title{Heavy quark potential at finite imaginary chemical potential}

\ShortTitle{Heavy quark potential at finite imaginary chemical potential}

\author{\speaker{Junichi Takahashi}, Takahiro Sasaki, Masanobu Yahiro\\
        Department of Physics, Graduate School of Sciences, Kyushu University,
             Fukuoka 812-8581, Japan\\
        E-mail: \email{takahashi@phys.kyushu-u.ac.jp}}
\author{Keitaro Nagata\\
        KEK Theory Center, High Energy Accelerator Research Organization (KEK),
             Tsukuba 305-0801, Japan
}
\author{Takuya Saito\\
        Integrated Information Center, Kochi University,
             Kochi 780-8520, Japan
}
\author{Hiroaki Kouno\\
        Department of Physics, Saga University,
             Saga 840-8502, Japan
}
\author{Atsushi Nakamura\\
        Research Institute for Information Science and Education, Hiroshima University,
             Higashi-Hiroshima 739-8527, Japan
}

\abstract{We investigate chemical-potential ($\mu$) dependence of static-quark free energies in both the real and imaginary $\mu$ regions, using the clover-improved two-flavor Wilson fermion action and the renormalization-group improved Iwasaki gauge action.
Static-quark potentials are evaluated from the Polyakov-loop correlator in the deconfinement phase and the imaginary $\mu=i\mu_{\rm I}$ region and extrapolated to the real $\mu$ region with analytic continuation.
As the analytic continuation, the potential calculated at imaginary $\mu=i\mu_{\rm I}$ is expanded into a Taylor-expansion series of  $i\mu_{\rm I}/T$ up to 4th order and the pure imaginary variable $i\mu_{\rm I}/T$ is replaced by the real one $\mu_{\rm R}/T$.
At real $\mu$, the 4th-order term weakens $\mu$ dependence of the potential sizably.
Also, the color-Debye screening mass is extracted from the color-singlet potential at imaginary $\mu$, and the mass is extrapolated to real $\mu$ by analytic continuation. 
The screening mass thus obtained has stronger $\mu$ dependence than the prediction of the leading-order thermal perturbation theory at both real and imaginary $\mu$.
This talk is based on \cite{Takahashi}.}

\FullConference{31st International Symposium on Lattice Field Theory - LATTICE 2013\\
		July 29 - August 3, 2013\\
		Mainz, Germany}

\begin{document}

\section{Introduction}
The free energies between two static quarks are fundamental quantities to understand inter-quark interactions.
Particularly above $T_{\rm pc}$, the static-quark potentials determined from the free energies characterize quark-gluon dynamics in QGP; for example, the inverse of the range of the color-singlet potential is the color-Debye screening mass.
The potential largely affects the behavior of heavy-quark bound states such as $J/\Psi$ and $\Upsilon$ in QGP created at the center of heavy-ion collisions~\cite{Matsui}.
In lattice QCD(LQCD) simulations, the static-quark potential is determined from the Polyakov-loop correlation function.
For zero chemical potential, $T$ dependence of the static-quark potential was investigated by quenched QCD~\cite{Kaczmarek1,Saito1,Saito2} and full QCD with staggered-type~\cite{Kaczmarek2} and Wilson-type quark actions~\cite{Bornyakov,Maezawa1,Maezawa2}. 
For small $\mu/T$, it was analyzed by the Taylor-expansion method with staggered-type~\cite{Doering} and Wilson-type quark actions~\cite{Ejiri}. 
In the analysis~\cite{Ejiri}, the expansion coefficients are taken up to 2nd order of $\mu/T$.\\
\indent
In this report, we present $\mu$ dependence of the static-quark free energies and the color-Debye screening mass in both the imaginary and real $\mu$ regions, performing LQCD simulations at imaginary $\mu$ and extrapolating the result to the real  $\mu=\mu_{\rm R}$ region with analytic continuation.
We consider two temperatures above $T_{\rm pc}$, i.e., $T/T_{\rm pc}=1.20$ and 1.35. 
Following the previous LQCD simulation~\cite{Ejiri} at small $\mu/T$, we compute static-quark free energies along the line of constant physics at $m_{\rm PS}/m_{\rm V}=0.80$. 
As the analytic continuation, the static-quark potential at imaginary $\mu=i\mu_{\rm I}$ is expanded into a Taylor-expansion series of $i\mu_{\rm I}/T$ and pure imaginary variable $i\mu_{\rm I}/T$ is replaced by real one $\mu_{\rm R}/T$.
In the present work the Taylor-expansion coefficients of the static-quark potential are evaluated up to 4th order, whereas the coefficients were computed up to 2nd order in Ref.~\cite{Ejiri}. 
It is found that the 4th-order term yields non-negligible contributions to $\mu$ dependence of the static-quark potentials at real $\mu$.
At long distance, all of the color singlet and non-singlet potentials tend to twice the single-quark free energy, indicating that the interactions between heavy quarks are fully color-screened.
Although this property is known for finite $T$ and zero $\mu$~\cite{Maezawa1}, the present work shows that the property persists also for imaginary $\mu$.
For imaginary $\mu$, the color-singlet $q{\bar q}$ and the color-antitriplet $qq$ interaction are attractive, whereas the color-octet $q{\bar q}$ and the color-sextet $qq$ interaction are repulsive. 
The color-Debye screening mass at imaginary $\mu$ is extracted from the color-singlet potential there.
The mass at real $\mu$ is extrapolated from the mass at imaginary $\mu$ by analytic continuation, i.e., by expanding the mass at imaginary $\mu$ into a power series of $i\mu_{\rm I}/T$ up to 2nd order and replacing $i\mu_{\rm I}$ by $\mu_{\rm R}$.
The $(\mu /T)$ dependence of the screening mass is found to be stronger than the prediction of the leading-order thermal perturbation theory.

\section{Static-quark free energies}
The Polyakov loop is defined as 
\begin{eqnarray}
L(\bm{x})=\prod^{N_{t}}_{t=1}U_{4}(\bm{x},t)
\end{eqnarray}
with link variables $U_{\mu} \in \mathrm{SU}(3)$. 
At imaginary $\mu$, the ensemble average of the Polyakov loop becomes a complex number, 
$\langle \mathrm{Tr}L(0) \rangle \equiv \Phi e^{i\theta}$. 
The modulus is related to the single-quark free energy $F_q$ as $ \Phi=\exp[-F_q/T]$. 
After an appropriate gauge fixing, one can derive 
the static-quark free energies (potentials)$V_M$ of color channel $M$ 
from the Polyakov-loop correlator~\cite{Nadkarni1,Nadkarni2}: 
\begin{eqnarray}
e^{-V_{1}(r,T,\mu)/T}&=&\frac{1}{3}\langle \mathrm{Tr}L^{\dag}(\bm{x})L(\bm{y}) \rangle, \\
e^{-V_{8}(r,T,\mu)/T}&=&\frac{1}{8}\langle \mathrm{Tr}L^{\dag}(\bm{x})\mathrm{Tr}L(\bm{y}) \rangle -\frac{1}{24}\langle \mathrm{Tr}L^{\dag}(\bm{x})L(\bm{y}) \rangle, \\
e^{-V_{3^{\ast}}(r,T,\mu)/T}&=&\frac{1}{6}\langle \mathrm{Tr}L(\bm{x})\mathrm{Tr}L(\bm{y}) \rangle -\frac{1}{6}\langle \mathrm{Tr}L(\bm{x})L(\bm{y}) \rangle,\\
e^{-V_{6}(r,T,\mu)/T}&=&\frac{1}{12}\langle \mathrm{Tr}L(\bm{x})\mathrm{Tr}L(\bm{y}) \rangle +\frac{1}{12}\langle \mathrm{Tr}L(\bm{x})L(\bm{y}) \rangle,
\end{eqnarray}
where $r=|\bm{x}-\bm{y}|$ and the subscripts $M=(1, 8, \mathrm{3}^{\ast}, 6)$ 
mean the color-singlet, -octet, -antitriplet and -sextet channels, 
respectively. 
We adopt the Coulomb gauge fixing.\\
\indent
In general, the $V_{M}$  ($M=1, 8, 3^{\ast}, 6$) are complex at finite imaginary $\mu$. 
The real part of $V_{M}$ is ${\cal C}$-even and the imaginary part is ${\cal C}$-odd. 
This can be easily understood by expanding $V_{M}$ into a power series of $i\mu_{\rm I}/T$:
\begin{eqnarray}
\frac{V_{M}(r,T,\mu_{\rm I})}{T}&=&v_{0}(r)+iv_{1}(r)\left(\frac{\mu_{\rm I}}{T}\right)+v_{2}(r)\left(\frac{\mu_{\rm I}}{T}\right)^{2}+iv_{3}(r)\left(\frac{\mu_{\rm I}}{T}\right)^{3}+v_{4}(r)\left(\frac{\mu_{\rm I}}{T}\right)^{4},  
\label{Eq:expansion-1}
\end{eqnarray}
where we consider terms up to 4th order. 
The potential $V_{M}$ at real $\mu$ is obtained from that at imaginary  $\mu$ by analytic continuation, i.e., by replacing $i\mu_{\rm I}/T$ by $\mu_{\rm R}/T$:
\begin{eqnarray}
\frac{V_{M}(r,T,\mu_{\rm R})}{T}&=&v_{0}(r)+v_{1}(r)\left(\frac{\mu_{\rm R}}{T}\right)-v_{2}(r)\left(\frac{\mu_{\rm R}}{T}\right)^{2}-v_{3}(r)\left(\frac{\mu_{\rm R}}{T}\right)^{3}+v_{4}(r)\left(\frac{\mu_{\rm R}}{T}\right)^{4}.  
\end{eqnarray}
The WHOT-QCD Collaboration calculated the Taylor-expansion coefficients of $V_{M}$ up to 2nd order by using the Taylor-expansion method and the reweighting technique with the Gaussian approximation for the distribution of the complex phase of the quark determinant~\cite{Ejiri}. 
In this work, meanwhile, we obtain the coefficients up to 4th order from $V_{M}$ at imaginary $\mu$ by expanding it as in (\ref{Eq:expansion-1}).

\section{Results of the lattice simulations and the analytic continuation}
\begin{figure}[htbp]
\begin{center}
\hspace{20pt}
  \includegraphics[width=0.45\textwidth]{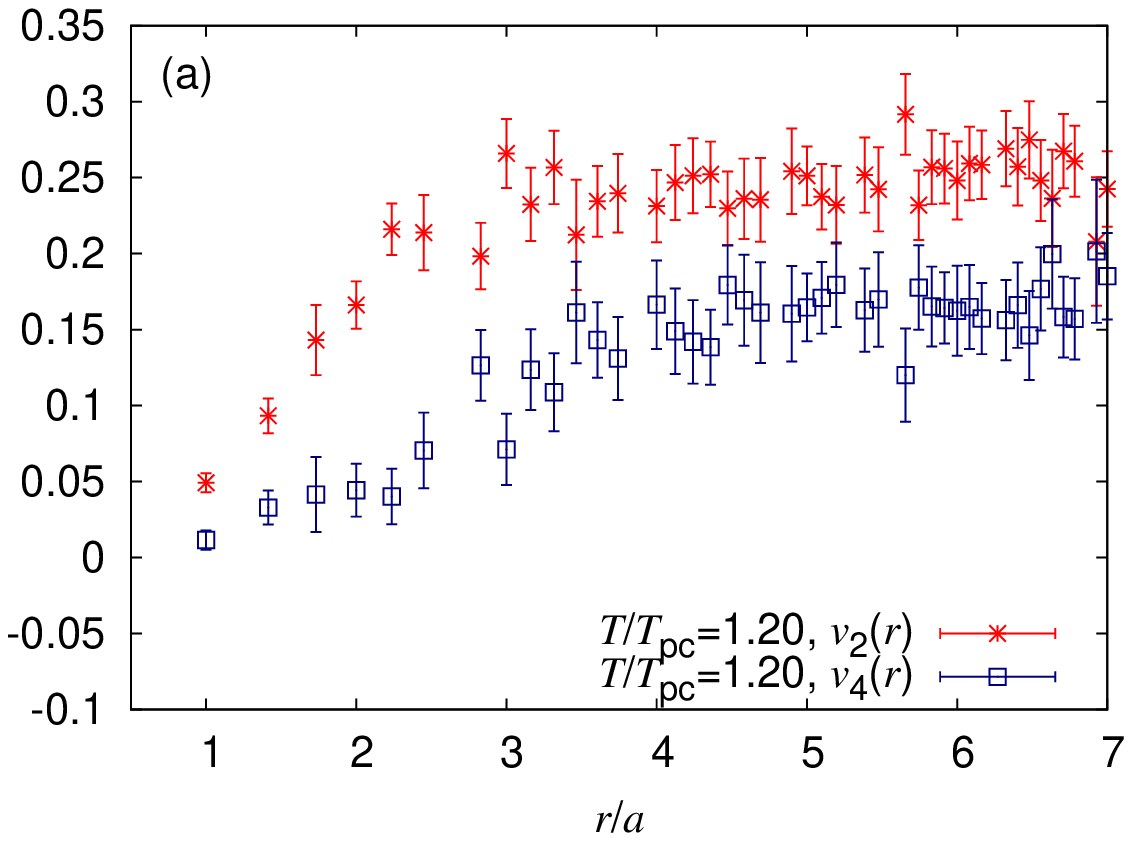}
  \includegraphics[width=0.45\textwidth]{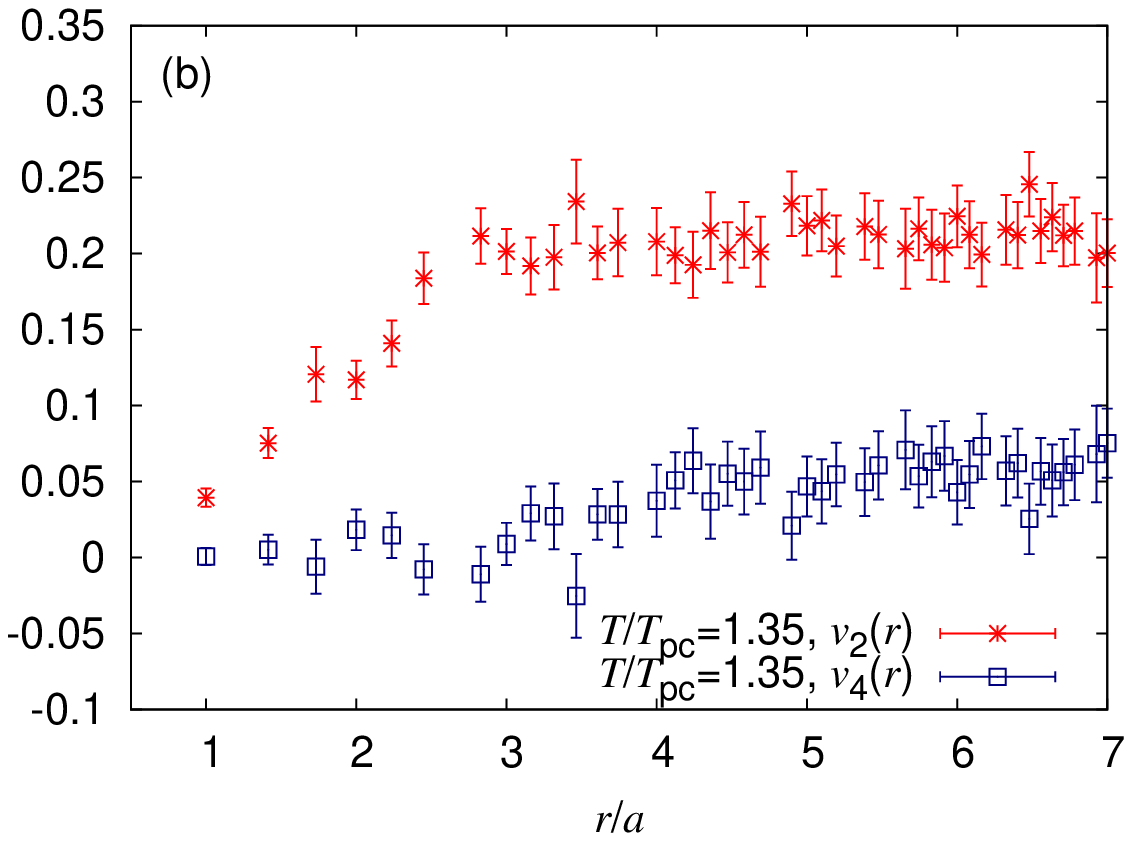}
\end{center}
\vspace{10pt}
\caption{Taylor-expansion coefficients, $v_2(r)$ and $v_4(r)$, of 
$V_{1}(r)$ at (a) $T/T_{\mathrm{pc}}=1.20$ and (b) $T/T_{\mathrm{pc}}=1.35$.
}
\label{Taylor-expansion-coefficient}
\end{figure}

We employ the clover-improved two-flavor Wilson fermion action and the renormalization-group improved Iwasaki gauge action.
Finite temperature simulations are performed on $16^{3} \times 4$ lattices along the line of constant physics with $m_{\mathrm{PS}}/m_{\mathrm{V}} = 0.80$.
We consider two temperatures $T/T_{\rm pc}=1.20$ and 1.35.
We generated 16,000 trajectories and removed the first 1,000 trajectories as thermalization for all the parameter set.
We measured the static-quark potential at every 100 trajectories.\\
\indent
The coefficients $v_2(r)$ and $v_4(r)$ of $V_{1}(r)$ are shown in Fig.~\ref{Taylor-expansion-coefficient}.
The ratio $v_4(r)/v_2(r)$ is about 3/4 for $T/T_{\mathrm{pc}}=1.20$ and about 1/4 for $T/T_{\mathrm{pc}}=1.35$.
Thus the contribution of $v_4(r)$ to $V_{1}(r)$ is significant near $T_{\mathrm{pc}}$ such as $T/T_{\mathrm{pc}}=1.20$. 
Even at higher $T$ such as $T/T_{\mathrm{pc}}=1.35$, the contribution is not negligible.

\begin{figure}[htbp]
\begin{center}
\hspace{20pt}
 \includegraphics[width=0.45\textwidth]{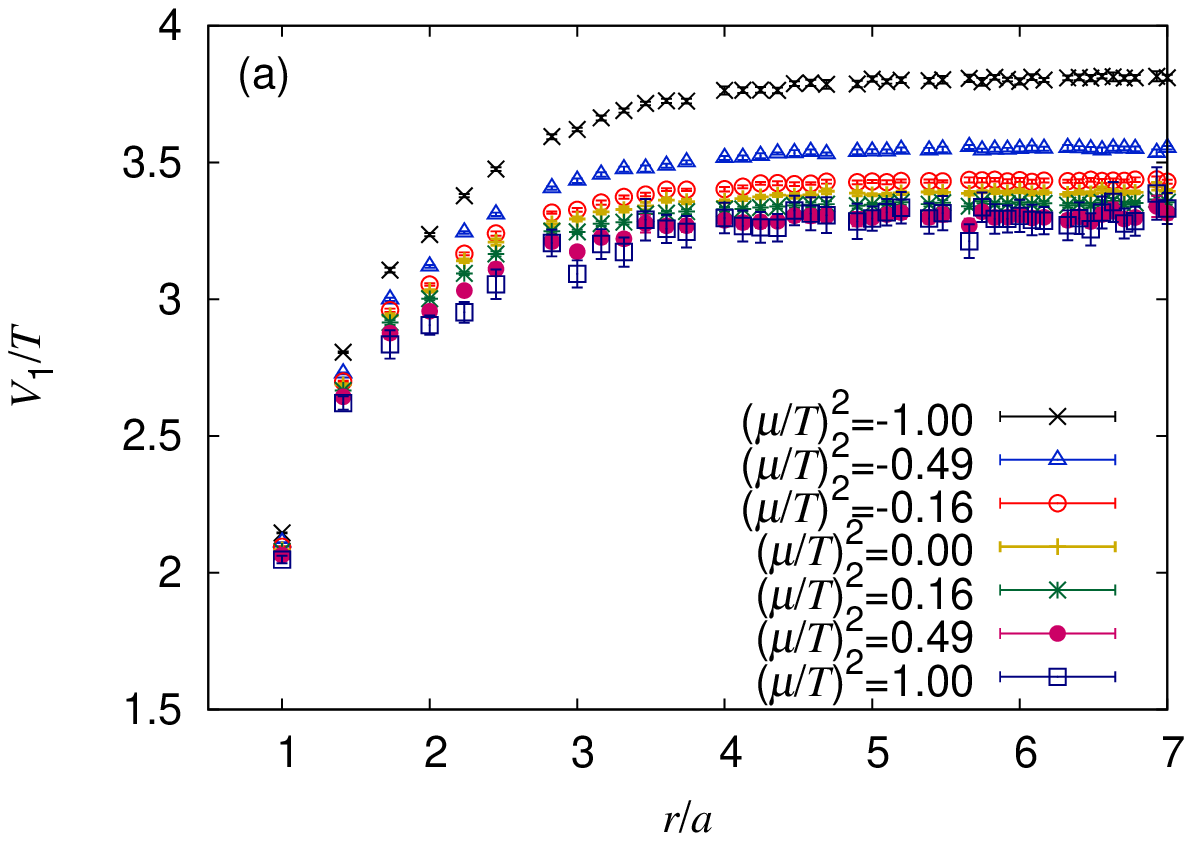}
 \includegraphics[width=0.45\textwidth]{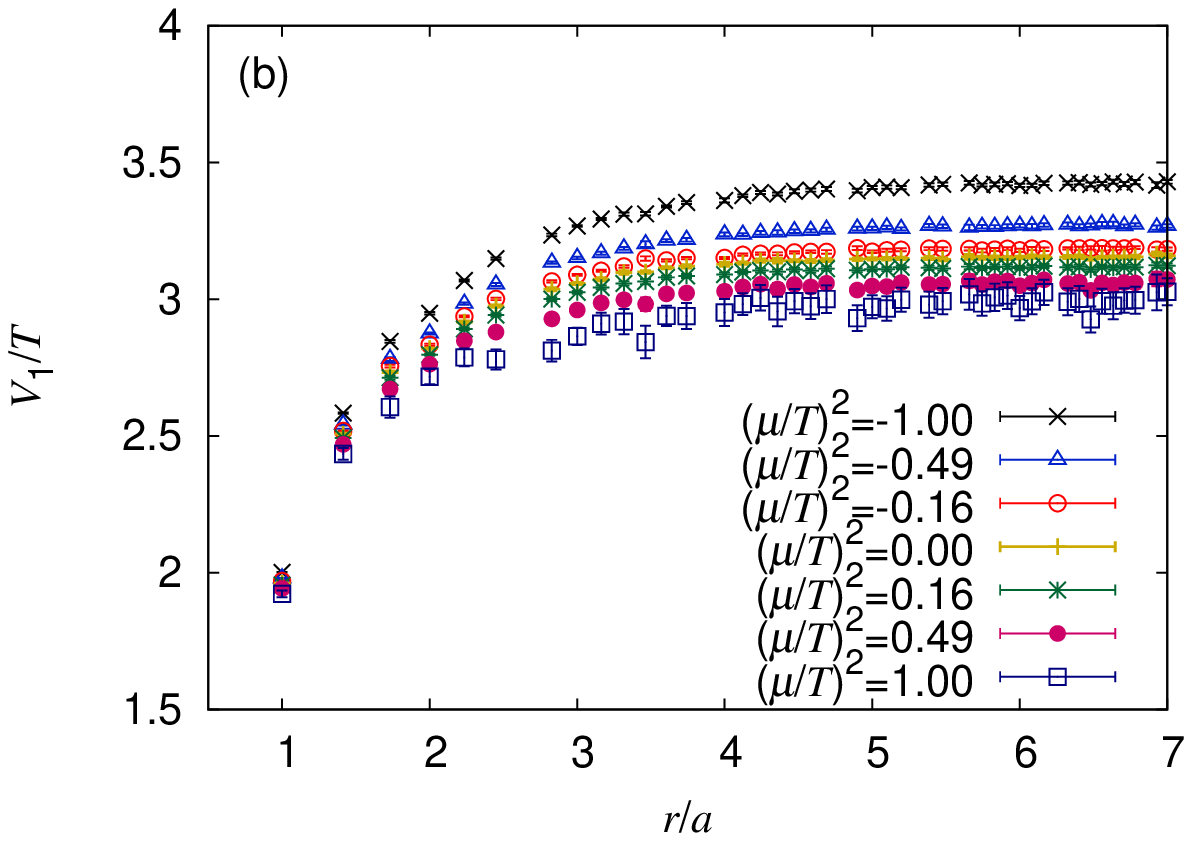}
\end{center}
\vspace{10pt}
\caption{
$\mu/T$ dependence of the color-singlet $q\bar{q}$ potential for (a) $T/T_{\mathrm{pc}}=1.20$ and (b) $T/T_{\mathrm{pc}}=1.35$.
}
\label{1pot_mu2-dep}
\end{figure}

\indent
Figure~\ref{1pot_mu2-dep} shows the color-singlet potential at imaginary and real $\mu$ for (a) $T/T_{\mathrm{pc}}=1.20$ and (b) $T/T_{\mathrm{pc}}=1.35$.
The chemical potential is varied from $(\mu/T)^{2}=-1.0$ to 1.0.
The potential $V_1$ is ${\cal C}$-even, so that  $v_{1}(r)=v_{3}(r)=0$. 
Furthermore, if $v_{4}(r)=0$, the potential $V_1/T$ will linearly depend on $(\mu/T)^{2}$. 
For $T/T_{\mathrm{pc}}=1.20$, $v_{4}(r)$ is comparable to $v_{2}(r)$.
For this property, in panel (a) of Fig.~\ref{1pot_mu2-dep}, $\mu/T$ dependence of  $V_1/T$ is much weaker at real $\mu$ than at imaginary $\mu$. 
In panel (b) of  $T/T_{\mathrm{pc}}=1.35$, $v_{4}(r)$ is still non-negligible compared with $v_{2}(r)$, so that $V_1/T$ has still weaker $\mu/T$ dependence at real $\mu$ than at imaginary $\mu$.

\begin{figure}[htbp]
\begin{center}
\hspace{20pt}
 \includegraphics[width=0.45\textwidth]{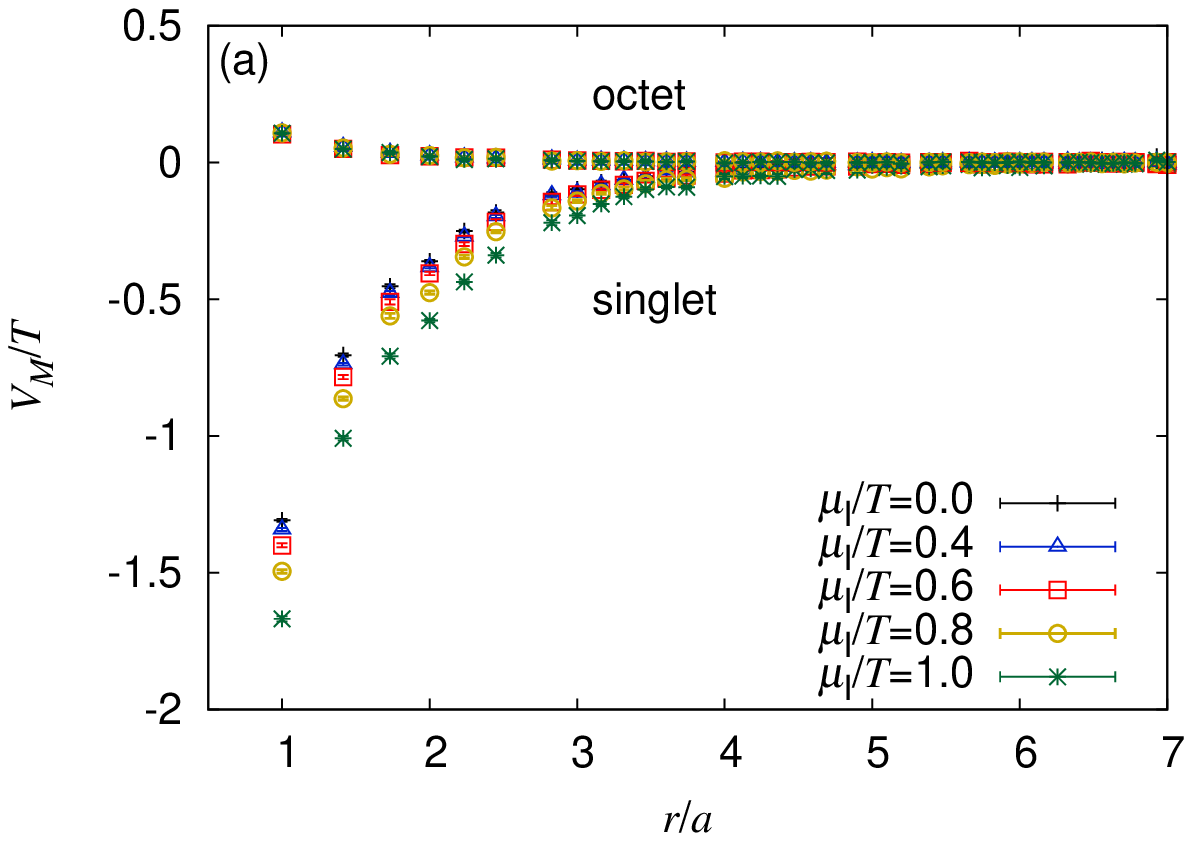}
 \includegraphics[width=0.45\textwidth]{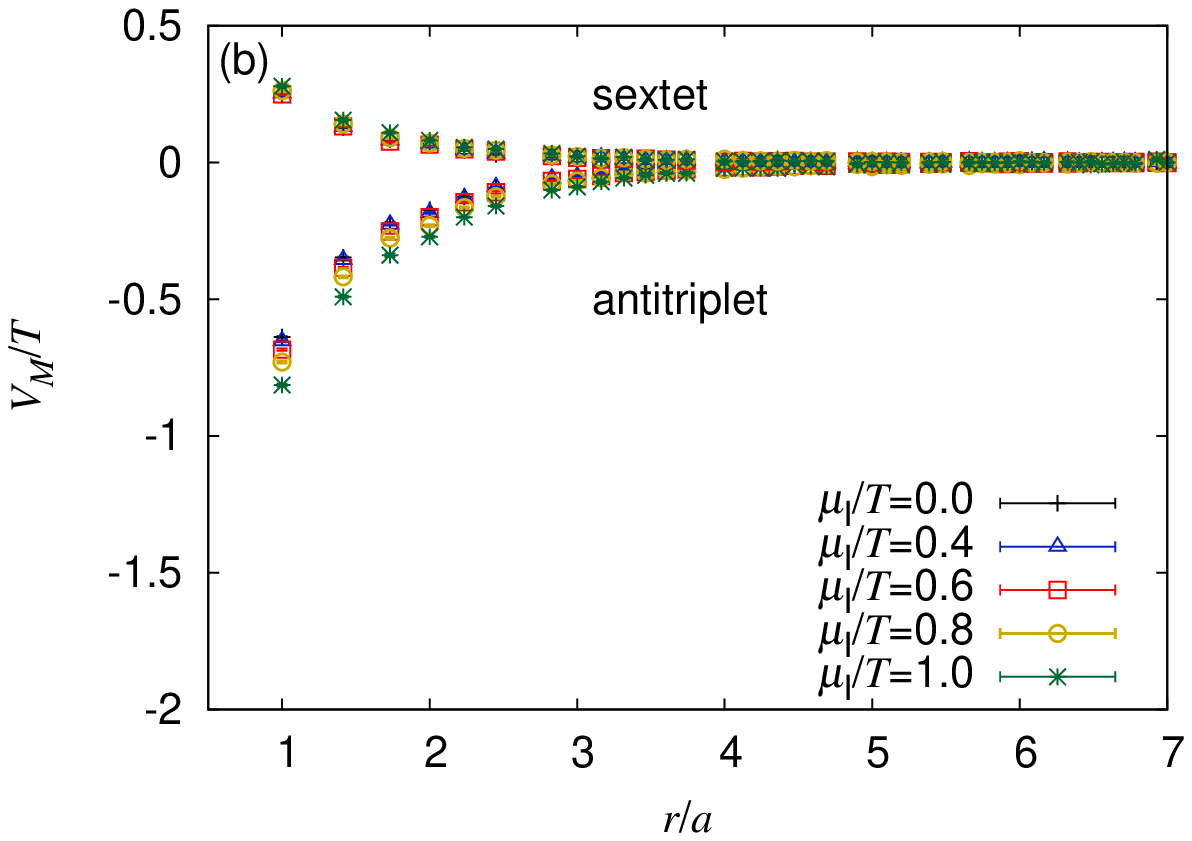}
\end{center}
\vspace{10pt}
\caption{
$\mu_{\rm I}/T$ dependence of the subtracted $q\bar{q}$ in (a) the color-singlet and -octet channels and $qq$ potentials in (b) the color-antitriplet and -sextet channels at $T/T_{\mathrm{pc}}=1.20$.
}
\label{pot_mu-dep_b1950_re}
\end{figure}

\indent
For the case of $T > T_{\rm pc}$ and $\mu=0$, the potentials $V_{M}(r)$ are known to tend to twice the single-quark free energy $2F_{q}$ in the limit of large $r$~\cite{Maezawa1}. 
This behavior persists also for imaginary $\mu$.
The interactions between heavy quarks are thus color screened also for imaginary $\mu$.
Following the previous works~\cite{Maezawa1,Saito1,Saito2,Maezawa2,Ejiri}, we then subtract $2F_{q}$ from $V_{M}(r)$.
The subtracted static-quark potentials are shown in Fig.~\ref{pot_mu-dep_b1950_re}(a) for the color-singlet and -octet channels and in Fig.~\ref{pot_mu-dep_b1950_re}(b) for the color-antitriplet and -sextet channels.
Needless to say, the physical interpretation of gauge dependent quantities is not straightforward; this is the case also for the potentials.
See ref.~\cite{Jahn:2004qr}.
Our results show distinctively different behaviors for the singlet/antitriplet channel  and the octet/sextet channel; the former is "attractive" and the latter is "repulsive".
The attractive interactions have strong $\mu_{\rm I}/T$ dependence, but the repulsive interactions have weak $\mu_{\rm I}/T$ dependence.

\section{Color-Debye screening mass}
\begin{figure}[htbp]
\begin{center}
\hspace{20pt}
 \includegraphics[width=0.45\textwidth]{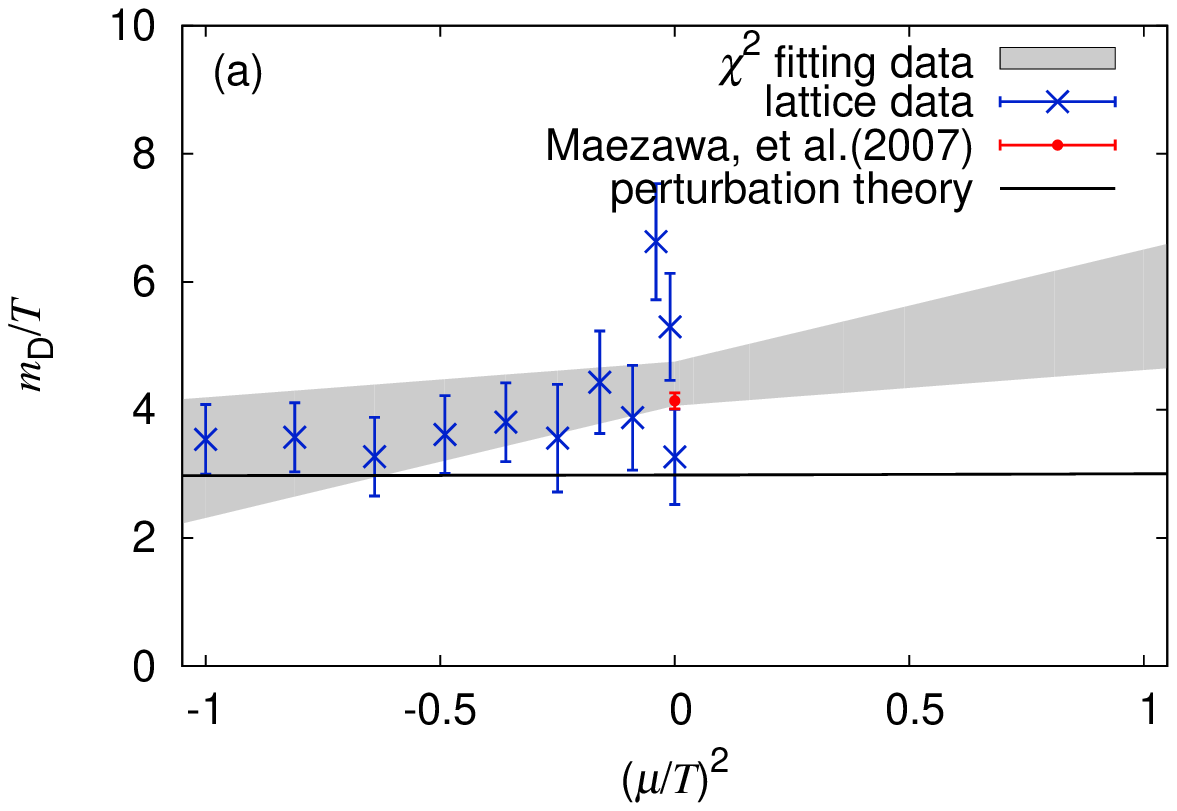}
 \includegraphics[width=0.45\textwidth]{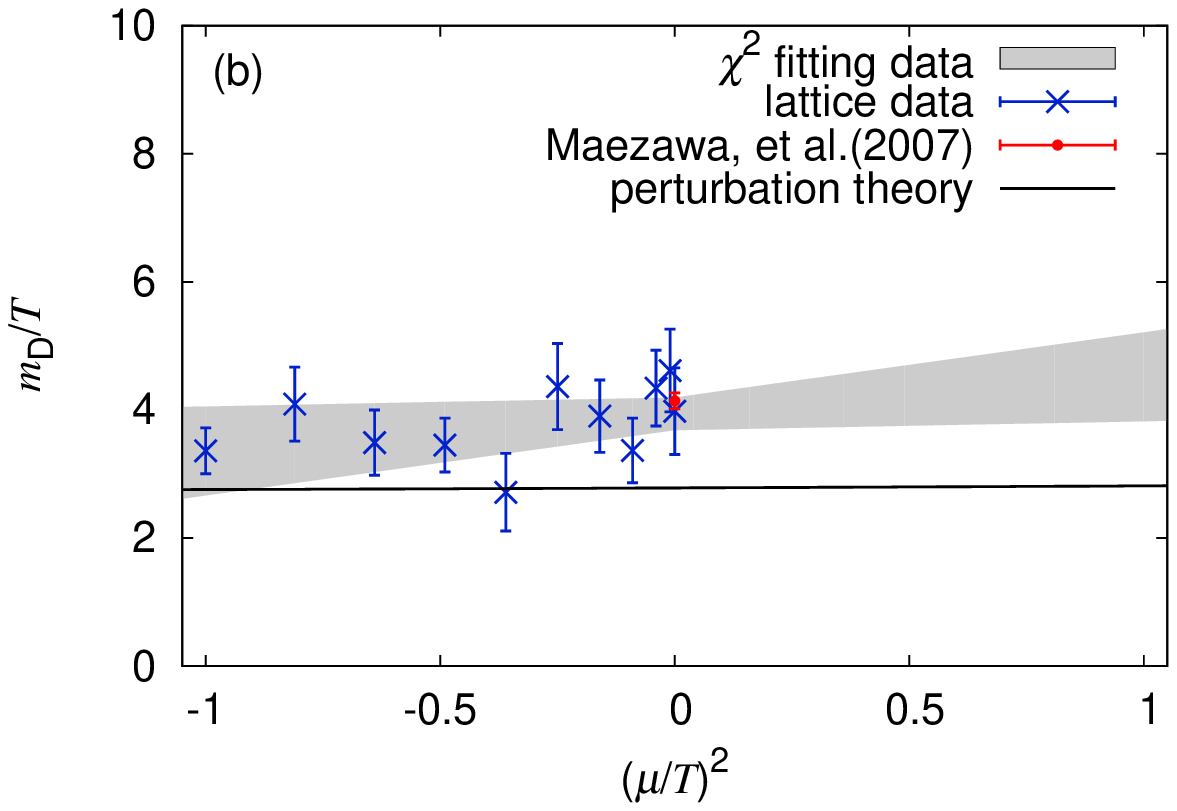}
\end{center}
\vspace{10pt}
\caption{
$(\mu/T)^{2}$ dependence of the color-Debye screening mass for (a) $T/T_{\mathrm{pc}}=1.20$ and (b) $1.35$.
The screening mass is determined from the singlet potential. 
Crosses with error bars denote results of the present lattice simulations at imaginary $\mu$, while a circle with an error bar is a result of the previous lattice simulations at $\mu=0$~\cite{Maezawa2}.
}
\label{mD_mu2dep}
\end{figure}

In order to analyze the color screening effect, we fit the static-quark potential to the screened Coulomb form
\begin{eqnarray}
V_{M}(r,T,\mu)=C_{M}\frac{\alpha_{\mathrm{eff}}(T,\mu)}{r}e^{-m_{\rm D}(T,\mu)r},
\label{eq:Yukawa-form}
\end{eqnarray}
where $C_{M}\equiv \langle \sum^{8}_{a=1}t^{a}_{1}\cdot t^{a}_{2}\rangle_{M}$, $\alpha_{\mathrm{eff}}$ and $m_{\rm D}(T,\mu)$ are the Casimir factor, the effective running coupling and the color-Debye screening mass, respectively. 
Here, we focus our discussion on the color-singlet channel that is most important in the real world, and the Casimir factor in the singlet channel is $C_{1}=-4/3$. 
Since $V_{1}=0$ in the limit of large $r$ in (\ref{eq:Yukawa-form}), we extract the screening mass from the subtracted static-quark potential.
Following the previous work~\cite{Maezawa2}, we choose a fit range of $\sqrt{11} \le r/a \le 6.0$.\\
\indent
In the leading-order (LO) hard thermal loop (HTL) perturbation theory, the color-Debye screening mass is obtained \cite{LeBellac} by
\begin{eqnarray}
\frac{m_{\rm D}(T,\mu)}{T}=g_{\mathrm{2l}}(\nu)\sqrt{\left(1+\frac{N_{f}}{6}\right)+\frac{N_{f}}{2\pi^{2}}\left(\frac{\mu}{T}\right)^{2}}
\end{eqnarray}
with the 2-loop running coupling $g_{\mathrm{2l}}$ given by
\begin{eqnarray}
g^{-2}_{\mathrm{2l}}(\nu)=\beta_{0}\ln\left(\frac{\nu}{\Lambda}\right)^{2}+\frac{\beta_{1}}{\beta_{0}}\ln \ln\left(\frac{\nu}{\Lambda}\right)^{2}, 
\end{eqnarray}
where the argument in the logarithms is rewritten into $\nu/\Lambda=(\nu/T)(T/T_{\mathrm{pc}})(T_{\mathrm{pc}}/\Lambda)$  with $\Lambda=\Lambda^{N_{f}=2}_{\bar{\mathrm{MS}}}\simeq 261$ MeV~\cite{Goeckeler} and $T_{\mathrm{pc}}\simeq 171$ MeV~\cite{Khan1}, and the renormalization point $\nu$ is assumed to be $\nu=\sqrt{(\pi T)^{2}+\mu^{2}}$~\cite{Ipp}.\\
\indent
Figure~\ref{mD_mu2dep} shows the $(\mu/T)^{2}$ dependence of the color-Debye screening mass for (a) $T/T_{\mathrm{pc}}=1.20$ and (b) $T/T_{\mathrm{pc}}=1.35$. 
The lattice-simulation results are plotted by the cross symbols.
The screening mass is then expanded up to 2nd order of $\mu/T$: 
\begin{eqnarray}
 \frac{m_{\rm D}}{T}=a_0(T) + a_2(T)\left(\frac{\mu}{T}\right)^{2},
\label{m_D-expansion}  
\end{eqnarray}
where note that $m_{\rm D}$ is ${\cal C}$-even and hence it has no linear term of $\mu/T$.
The screening mass at real $\mu$ is extrapolated from that at imaginary $\mu$ by using (\ref{m_D-expansion}).

The results of the extrapolation, represented by the hatching area, are consistent with the previous LQCD result, denoted by a circle symbol, at $\mu=0$~\cite{Maezawa2} for both cases of $T/T_{\mathrm{pc}}=1.20$ and  $1.35$.
Comparing the hatching area (the result of the extrapolation) with the solid line (the prediction of the leading-order thermal perturbation theory), one can see that the present LQCD results show stronger $\mu/T$ dependence than the prediction of the perturbation theory.

\section{Summary}
We have investigated $\mu$ dependence of the static-quark potential and the color-Debye screening mass in both the imaginary and real $\mu$ regions, performing LQCD simulations at imaginary $\mu$ and extrapolating the result to the real $\mu$ region with analytic continuation.
LQCD calculations are done on a $16^{3}\times 4$ lattice with the clover-improved two-flavor Wilson fermion action and the renormalization-group improved Iwasaki gauge action.
We took an intermediate quark mass and considered two cases of $T/T_{\mathrm{pc}}=1.20$ and $1.35$.\\
\indent
The static-quark potential at real $\mu$ was obtained by expanding the potential at imaginary $\mu$ into a Taylor-expansion series of $i\mu_{\rm I}/T$ up to 4th order and replacing $i\mu_{\rm I}$ to $\mu_{\rm R}$.
Since the expansion series was taken only up to 2nd order in the previous analysis~\cite{Ejiri}, this is the first analysis that investigates contributions of the 4th-order term to the potential.
We found that at real $\mu$ the 4th-order term weakens $\mu$ dependence of the potential sizably.
This effect becomes more significant as $T$ decreases toward $T_{\mathrm{pc}}$.
We have also investigated color-channel dependence of the static-quark potentials.
At large distance, all the potentials tend to twice the single-quark free energy, indicating that the interactions are fully color screened.
Although this property is known for finite $T$ and zero $\mu$~\cite{Maezawa1}, the present analysis shows that the property persists also for imaginary $\mu$.
For imaginary $\mu$, the color-singlet $q{\bar q}$ and the color-antitriplet $qq$ interaction are attractive, whereas the color-octet $q{\bar q}$ and the color-sextet $qq$ interaction are repulsive.\\
\indent
The color-Debye screening mass is evaluated from the color-singlet potential at imaginary $\mu$.
The screening mass thus obtained at imaginary $\mu$ is extrapolated to real $\mu$ by expanding the mass at imaginary $\mu$ into a power series of $i\mu_{\rm I}/T$ up to 2nd order and replacing $i\mu_{\rm I}/T$ by $\mu_{\rm R}/T$.
The resulting mass has stronger $\mu$ dependence at both imaginary and  real $\mu$ than the prediction of the leading-order thermal perturbation theory.

\noindent
\begin{acknowledgments}
Junichi Takahashi is supported by JSPS KAKENHI (No. 25-3944), Takahiro Sasaki by JSPS KAKENHI (No. 23-2790),
Atsushi Nakamura by JSPS KAKENHI (Nos. 23654092, 24340054) and Takuya Saito
by JSPS KAKENHI (No. 23740194). Keitaro Nagata is supported in part by Strategic Programs for Innovative Research (SPIRE) Field 5. The numerical calculations were performed on NEC SX-9 and SX-8R at CMC, Osaka University.\\
\end{acknowledgments}



\begin{thebibliography}{99}
\expandafter\ifx\csname natexlab\endcsname\relax\def\natexlab#1{#1}\fi
\expandafter\ifx\csname bibnamefont\endcsname\relax
  \def\bibnamefont#1{#1}\fi
\expandafter\ifx\csname bibfnamefont\endcsname\relax
  \def\bibfnamefont#1{#1}\fi
\expandafter\ifx\csname citenamefont\endcsname\relax
  \def\citenamefont#1{#1}\fi
\expandafter\ifx\csname url\endcsname\relax
  \def\url#1{\texttt{#1}}\fi
\expandafter\ifx\csname urlprefix\endcsname\relax\def\urlprefix{URL }\fi
\providecommand{\bibinfo}[2]{#2}
\providecommand{\eprint}[2][]{\url{#2}}
%
\bibitem{Takahashi}
\bibinfo{author}{\bibfnamefont{J.}~\bibnamefont{Takahashi et al.}}, 
\bibinfo{howpublished}{arXiv:1308.2489 [hep-lat]},
(\bibinfo{year}{2013}).
\bibitem{Matsui}
\bibinfo{author}{\bibfnamefont{T.}~\bibnamefont{Matsui}}
\bibnamefont{and}
\bibinfo{author}{\bibfnamefont{H.}~\bibnamefont{Satz}},  
\bibinfo{journal}{Phys. Lett. B} \textbf{\bibinfo{volume}{178}},
\bibinfo{pages}{416} (\bibinfo{year}{1986}). 
\bibitem{Kaczmarek1}
\bibinfo{author}{\bibfnamefont{O.}~\bibnamefont{Kaczmarek}},
\bibinfo{author}{\bibfnamefont{F.}~\bibnamefont{Karsch}},
\bibinfo{author}{\bibfnamefont{E.}~\bibnamefont{Laermann}},
\bibnamefont{and}
\bibinfo{author}{\bibfnamefont{M.}~\bibnamefont{Lutgemeier}},
\bibinfo{journal}{Phys. Rev. D} \textbf{\bibinfo{volume}{62}},
\bibinfo{pages}{034021} (\bibinfo{year}{2000}). 
\bibitem{Saito1}
\bibinfo{author}{\bibfnamefont{A.}~\bibnamefont{Nakamura}}, 
\bibnamefont{and}
\bibinfo{author}{\bibfnamefont{T.}~\bibnamefont{Saito}}, 
\bibinfo{journal}{Prog. Theor. Phys.} \textbf{\bibinfo{volume}{111}},
\bibinfo{pages}{733} (\bibinfo{year}{2004}).
\bibitem{Saito2}
\bibinfo{author}{\bibfnamefont{A.}~\bibnamefont{Nakamura}}, 
\bibnamefont{and}
\bibinfo{author}{\bibfnamefont{T.}~\bibnamefont{Saito}}, 
\bibinfo{journal}{Phys. Lett. B} \textbf{\bibinfo{volume}{621}}
\bibinfo{pages}{171} (\bibinfo{year}{2005}).
\bibitem{Kaczmarek2}
\bibinfo{author}{\bibfnamefont{O.}~\bibnamefont{Kaczmarek}},
\bibnamefont{and}
\bibinfo{author}{\bibfnamefont{F.}~\bibnamefont{Zantow}},
\bibinfo{journal}{Phys. Rev. D} \textbf{\bibinfo{volume}{71}},
\bibinfo{pages}{114510} (\bibinfo{year}{2005}). 
\bibitem{Bornyakov}
\bibinfo{author}{\bibfnamefont{V.}~\bibfnamefont{G.}~\bibnamefont{Bornyakov et al.~(DIK Collaboration)}}, 
\bibinfo{journal}{Phys. Rev. D} \textbf{\bibinfo{volume}{71}},
\bibinfo{pages}{114504} (\bibinfo{year}{2005}). 
\bibitem{Maezawa1}
\bibinfo{author}{\bibfnamefont{Y.}~\bibnamefont{Maezawa et al.~(WHOT-QCD Collaboration)}}, 
\bibinfo{howpublished}{arXiv:1112.2756 [hep-lat]},
(\bibinfo{year}{2012}).
\bibitem{Maezawa2}
\bibinfo{author}{\bibfnamefont{Y.}~\bibnamefont{Maezawa et al.~(WHOT-QCD Collaboration)}}, 
\bibinfo{journal}{Phys. Rev. D} \textbf{\bibinfo{volume}{75}},
\bibinfo{pages}{074501} (\bibinfo{year}{2007}). 
\bibitem{Doering}
\bibinfo{author}{\bibfnamefont{M.}~\bibnamefont{D\"{o}ring}},
\bibinfo{author}{\bibfnamefont{S.}~\bibnamefont{Ejiri}},
\bibinfo{author}{\bibfnamefont{O.}~\bibnamefont{Kaczmarek}},
\bibinfo{author}{\bibfnamefont{F.}~\bibnamefont{Karsch}},
\bibnamefont{and}
\bibinfo{author}{\bibfnamefont{E.}~\bibnamefont{Laermann}},
\bibinfo{journal}{Eur. Phys. J. C} \textbf{\bibinfo{volume}{46}},
\bibinfo{pages}{179} (\bibinfo{year}{2006}). 
\bibitem{Ejiri}
\bibinfo{author}{\bibfnamefont{S.}~\bibnamefont{Ejiri et al.~(WHOT-QCD Collaboration)}}, 
\bibinfo{journal}{Phys. Rev. D} \textbf{\bibinfo{volume}{82}},
\bibinfo{pages}{014508} (\bibinfo{year}{2010}). 
\bibitem{Nadkarni1}
\bibinfo{author}{\bibfnamefont{S.}~\bibnamefont{Nadkarni}},
\bibinfo{journal}{Phys. Rev. D} \textbf{\bibinfo{volume}{33}},
\bibinfo{pages}{3738} (\bibinfo{year}{1986}). 
\bibitem{Nadkarni2}
\bibinfo{author}{\bibfnamefont{S.}~\bibnamefont{Nadkarni}},
\bibinfo{journal}{Phys. Rev. D} \textbf{\bibinfo{volume}{34}},
\bibinfo{pages}{3904} (\bibinfo{year}{1986}). 
\bibitem{Jahn:2004qr}
\bibinfo{author}{\bibfnamefont{O.}~\bibnamefont{Jahn}},
\bibnamefont{and}
\bibinfo{author}{\bibfnamefont{O.}~\bibnamefont{Philipsen}},
\bibinfo{journal}{Phys. Rev. D} \textbf{\bibinfo{volume}{70}},
\bibinfo{pages}{074504} (\bibinfo{year}{2004}).
\bibitem{LeBellac}
\bibinfo{author}{\bibfnamefont{Michel}~\bibfnamefont{Le}~\bibnamefont{Bellac}},
{\it Thermal Field Theory}, Cambridge University Press, Cambridge
(\bibinfo{year}{1996}). 
\bibitem{Goeckeler}
\bibinfo{author}{\bibfnamefont{M.}~\bibnamefont{G\"{o}ckeler et al.}}, 
\bibinfo{journal}{Phys. Rev. D} \textbf{\bibinfo{volume}{73}},
\bibinfo{pages}{014513} (\bibinfo{year}{2006}). 
\bibitem{Khan1}
\bibinfo{author}{\bibfnamefont{A.}~\bibfnamefont{Ali}~\bibnamefont{Khan et al.~(CP-PACS Collaboration)}},
\bibinfo{journal}{Phys. Rev. D} \textbf{\bibinfo{volume}{63}},
\bibinfo{pages}{034502} (\bibinfo{year}{2000}). 
\bibitem{Ipp}
\bibinfo{author}{\bibfnamefont{A.}~\bibnamefont{Ipp}}, 
\bibnamefont{and}
\bibinfo{author}{\bibfnamefont{A.}~\bibnamefont{Rebhan}}, 
\bibinfo{journal}{J. High Energy Phys.} \textbf{\bibinfo{volume}{06}}
\bibinfo{pages}{032} (\bibinfo{year}{2003}).

\end{thebibliography}
\end{document}